\documentclass[pra,amsmath,amssymb,showpacs,superscriptaddress,twocolumn]{revtex4}
\usepackage{graphicx}
\usepackage{float}
\usepackage{amsthm,amsfonts}
\begin{document}
\title{Quantum fluctuation of entanglement for accelerated two-level detectors}

\author{Sixuan Zhang, Tonghua Liu\footnote{liutongh@mail.bnu.edu.cn}, Shuo Cao\footnote{caoshuo@bnu.edu.cn}, Yuting Liu, Shuaibo Geng, and Yujie Lian}
\affiliation{Department of Astronomy, Beijing Normal University,
Beijing 100875, China}

\begin{abstract}

\vspace*{0.2cm} Quantum entanglement as the one of the most general quantum resources, can be quantified by von Neumann entropy. However, as we know, the von Neumann entropy is only statistical quantity or operator, it therefore has fluctuation. The quantum fluctuation of entanglement (QFE) between Unruh-Dewitt detector modeled by a two-level atom is investigated in a relativistic setting.  The Unruh radiation and quantum fluctuation effects affect the precise measurement of quantum entanglement. Inspired by this we present how the relativistic motion effects QFE for two entangled Unruh-Dewitt detectors when one of them is accelerated
and interacts with the neighbor external scalar field. We find that QFE first increases by the Unruh thermal noise and then suddenly decays when the acceleration reaches at a considerably large value, which indicates that relativistic effect will lead to non-negligible QFE effect.  We also find that the initial QFE (without acceleration effect) is minimum with the maximally entangled state. Moreover, although QFE has a huge decay when the acceleration is greater than $\sim0.96$, concurrence also decays to a very low value, the ratio $\Delta E/C$ therefore still large. According to the equivalence principle, our  findings could be in principle applied to
dynamics of QFE under the influence of gravitation field.

 \end{abstract}

\maketitle
\section{Introduction.}
The difference between classical and quantum theory lies in the fact
that physical quantities in quantum mechanics are corresponded to
Hermitian operator, which could directly determine the state of the
quantum system \cite{T1A}. Thus, the eigenvalues are the possible
values of the Hermitian operator acting on the quantum state via
measurements. However, one should note that the outcome of each
measurement of a physical quantity shouldn't be identical. For a
physical quantity in quantum mechanics, the expected value is
usually used to represent the eigenvalue of the quantity, i.e., the
expected value stands for the statistical mean of the measurement
outcome by repeated several times measurements. As is known, the
fluctuations are inevitable in practical measurement processes. More
specifically, for a physical quantities $A$ in the quantum state
$\psi$, the fluctuation is defined as $\overline{\Delta
A^2}=\overline{(\hat{A}-\overline{A})^2}=\langle \psi
|(\hat{A}-\overline{A})^2|\psi\rangle$. Focusing on one of the most
general quantum resources, quantum entanglement, it can be
quantified by one statistical quantity or operator, von Neumann
entropy \cite{T1,T2}. Recently, the relations between the
fluctuation of quantified entropy and entanglement has been widely
studied in Refs. \cite{C1,C2,C3,C4}.

On the other hand, as one of the most important developments in
modern physics, quantum entanglement has been extensively
investigated in most recent years \cite{T3,T4,T5}. It is
interesting to note that, the importance of quantum entanglement not
only embodies the fundamental perspective of quantum information
task, but also attributes to its advantages in practical aspects
\cite{T7,T8,T9}. Although quantum entanglement has been achieved in
many experiments, however, most of these measurements were carried
out without considering the effect of acceleration. Actually, in
realistic situation, the preparation of quantum system and the
procession of quantum information tasks are always accompanied by
accelerated effects \cite{T9a1,T9a2,T9a3,T9a4,T9a,T9b}. In the framework of such accelerated quantum
system, the Unruh effect will be generated, which indicates that
quantum properties of fields are observer dependent \cite{T10,T11}.
From theoretical point of view, the Unruh effect will reveal thermal
radiation detected by a uniformly accelerated detector in the
Minkowski vacuum and that associated with the proper acceleration of
the detector. Following this direction, a number of combined
analyses involving the dynamics of quantum entanglement and steering
between two correlated Unruh-DeWitt detectors have been performed in
the literature \cite{T12,T5,T13}, which indicated that the type of
quantum resource will be reduced by the Unruh effect, while the
acceleration effect on quantum systems is non-negligible when using
quantum resources to perform quantum information task
\cite{T13A,T13B,T13A1,T13A2,T13A3}.

Inspire by above works, in this paper we will investigate the
quantum fluctuation of entanglement (QFE) for a two-level atom
accelerator, which is modeled by Unruh-DeWitt detectors in the
relativistic setting. Compared with the global free models
extensively used in many papers \cite{T14,T15}, the Unruh-DeWitt
detector model applied in this analysis to study the behavior of QFE
in a non-inertial system \cite{T16} has more advantages. On the one
hand, the problems related to single-mode approximation and
physically unfeasible detection of quantum correlations in the full
space-time can be effectively avoided \cite{T17}, which could
provide us a better understanding of quantum entanglement. On the
other hand, precise measurements of quantum entanglement are highly
dependent on the Unruh radiation and quantum fluctuation effects
\cite{T17A}, which supports a quantitative analysis of QFE in
relativistic setting. In this paper, we will carry out a
quantitative analysis and explore how the Unruh radiation affects
the QFE. This paper is organized as follows. In Sec. II, we briefly
describe the Unruh-DeWitt detectors and the evolution of the
prepared state under the Unruh thermal bath. In Sec. III we
introduce the quantum fluctuation of entanglement. The behaviors of
QFE with the detectors model in a relativistic setting are are
presented in Sec. IV. Finally, we summarize our conclusions in Sec.
V.

\section{Evolution of the detectors' state under relativistic motion}

In this section, from the view point of quantum information, we will
give a brief description of the Unruh-DeWitt detectors \cite{T12}
and furthermore discuss the dynamics of a pair of detectors
(considering the relativistic motion of one detector). Based on the
well known two-level atom system, the simplest model for one quantum
system, our Unruh-DeWitt detectors are modeled by a point-like
two-level atoms (each atom represents a detector). However, in our
analysis we also try to extend the interaction term and investigate
the interaction between this two-level system and its nearby fields.
Note that the detector is semiclassical, because it possesses a
classical world line but its internal degree of freedom is treated
quantum mechanically. In order to investigate the behavior of
quantum properties, one usually assume that the detectors are
initially sharing some quantum correlations between the Minkowski
spacetime and observed by two observers called Alice and Rob,
respectively. Alice and Rob's detectors initially are prepared for
the inertial frame, then we let that Alice still keeps inertial and
always be switched off, while Rob's detector interacts with the
scalar field, and moves with uniform acceleration for a time
duration $\Delta$. The Rob's detector will move with its world line
as \begin{eqnarray}\label{worldline} t(\tau)&=&a^{-1}\sinh a\tau,\; \\
\nonumber x(\tau)&=&a^{-1}\cosh a\tau,\; y(\tau)=z(\tau)=0,\;
\end{eqnarray}
where $a$ is the detector's proper acceleration, $\tau$ represents
the detector's proper time and $(t,x,y,z)$ are the usual Cartesian
coordinates in the Minkowski spacetime. For convenience, we employ
the natural units $c=\hbar=\kappa_{B}=1$ throughout this paper.

Considering the interaction between Rob's detector and the field,
the initial state of the total system can be expressed as
\begin{equation}
|\Psi_{t_0}^{AR\phi}\rangle=|\Psi_{AR}\rangle\otimes|0_{M}\rangle
,\label{IS}%
\end{equation}
where $|\Psi_{AR}\rangle=\sin \theta
|0_{A}\rangle|1_{R}\rangle+\cos\theta|1_{A}\rangle |0_{R}\rangle$
represents the initial state shared by Alice's (A) and Rob's (R)
detectors, $\theta$ denotes the initial state parameter, and
$|0_{M}\rangle$ represents the contribution of the external scalar
field in Minkowski vacuum. Now the total Hamiltonian of the system
with the scalar field can be written as
\begin{equation}
H_{A\, R\, \phi} = H_A + H_R + H_{KG} +  H^{R\phi}_{\rm int},\label{totalh}
\end{equation}
where $KG$ (Klein-Gordon) denotes the scalar field, while
$H_{A}=\Omega A^{\dagger}A$ and $H_{R}=\Omega R^{\dagger}R$ are
Hamiltonians for the creation and annihilation operators ($\Omega$
is the energy gap of the detectors). The interaction term,
$H^{R\phi}_{\rm int}(t)$, which describes how Rob's detector is
coupled with the external massless scalar field $\phi$, satisfies
the Klein-Gordon (KG) equation \cite{T18}
\begin{equation}
H^{R\phi}_{\rm int}(t)=
\epsilon(t) \int_{\Sigma_t} d^3 {\bf x} \sqrt{-g} \phi(x) [\chi({\bf x})R +
                           \overline{\chi}({\bf x})R^{\dagger}],
\label{int}
\end{equation}
where $g\equiv {\rm det} (g_{ab})$, $g_{ab}$ is the Minkowski
metric, and $\bf{x}$ is the coordinates defined on the Cauchy
surface $\sum_t=const$, which is associated with the time-like
isometries followed by the qubits. Meanwhile, $\epsilon(t)$
represents a real-valued switching function, which keeps the
detector switched on smoothly for a finite amount of proper time
$\Delta$. A point-like coupling function,
$\chi(\mathbf{x})=(\kappa\sqrt{2\pi})^{-3}\exp(-\mathbf{x}^{2}/2\kappa^{2})$,
guarantees that the space-localized detector interacts only with the
field in a neighborhood of its world line. For simplicity, we take
the detector-field free Hamiltonian as $H_0 = H_A + H_R + H_{KG}$,
based on which the total Hamiltonian can be rewritten as
\begin{equation}
H_{AR\, \phi}= H_0 + H_{\rm int}.
\end{equation}

Now working with the interaction term, we turn the state into the
interaction representation labeled by $I$ and rewrite the final
state $|\Psi^{AR \phi}_t \rangle$ at the time of $t=t_0+\Delta$:
\begin{equation}
|\Psi^{AR \phi}_t \rangle =
T \exp[-i\int_{t_0}^t dt H_{\rm int} ^I(t)] |\Psi^{AR \phi}_{t_0} \rangle,
\label{Dyson1}
\end{equation}
where $T$ is the time-ordering operator, the term $H_{\rm int}^I (t)
= U^{\dagger}_0(t) H_{\rm int} (t) U_0 (t)$, and $U_0 (t)$ is the
unitary evolution operator associated with the Hamiltonian $H_0$.
The dynamics of the atom-field system at $t=t_0+\Delta$ can be
calculated by the first order of perturbation over the coupling
constant $\epsilon$ \cite{T12}. Based on the dynamic evolution
described by the Hamiltonian given by Eq.(\ref{totalh}), the final
state $|\Psi^{AR \phi}_{t} \rangle$ could be expressed as
\cite{T13,T16}
\begin{equation}
|\Psi^{AR \phi}_{t} \rangle
= [I - i(\phi(f)R + \phi(f)^{\dagger} R^{\dagger}) ] |\Psi^{AR \phi}_{t_0} \rangle,
\label{primeira_ordem}
\end{equation}
where $I$ is the identity matrix with the same dimension of
$|\Psi^{AR \phi}_{t_0} \rangle$, and the operator
\begin{eqnarray}
\nonumber\phi(f) &\equiv& \int d^4 x \sqrt{-g}\chi(x)f\\&=&i [a_{RI}(\overline{u E\overline{f}})-a_{RI}^{\dagger}(u Ef)],
\label{phi(f)}
\end{eqnarray}
is the distribution function corresponding to the external scalar
field. Here $Ef$ is approximately a positive-frequency solution of
the scalar field \cite{T13,T16}, while the operator $u$ is an
positive-frequency solution of the Klein-Gordon equation, with
respect to the time-like isometry. Therefore, combined with the
initial states [Eq.~(\ref{IS}) and Eq.~(\ref{primeira_ordem})], the
final state of the total system can be expressed in terms of the
Rindler operators ($a_{R I}^{\dagger}$ and $a_{R I}$), i.e.,
\begin{eqnarray}
| \Psi^{AR \phi}_{t}\rangle
& = &
|\Psi^{AR \phi}_{t_0} \rangle
 + \sin \theta |0_A\rangle  |0_R\rangle
 \otimes(a_{R I}^{\dagger}(\lambda)|0_M\rangle)
 \nonumber \\
& + & \cos \theta |1_A\rangle |1_R\rangle\otimes(a_{R I}(\overline{\lambda})|0_M\rangle),
\label{evolutionAUX}
\end{eqnarray}
where $\lambda = -uEf$, while the creation operator $a_{R
I}^{\dagger}(\lambda)$ and annihilation operator $a_{R
I}(\overline{\lambda)}$ are defined in the Rindler region $I$. It
should be pointed out that the relation between these two sets of
operators are \cite{T13,T16}
\begin{eqnarray}
a_{R I}(\overline{\lambda})&=&
\frac{a_M(\overline{F_{1 \Omega}})+
e^{-\pi \Omega/a} a_M ^{\dagger} (F_{2 \Omega})}{(1- e^{-2\pi\Omega/a})^{{1}/{2}}},
\label{aniq} \\
a^{\dagger}_{R I}(\lambda)&=&
\frac{a^{\dagger}_M (F_{1 \Omega}) +
e^{-\pi \Omega/a}a_M(\overline{F_{2 \Omega}})}{(1- e^{-2\pi\Omega/a})^{{1}/{2}}}
\label{cria},
\end{eqnarray}
with $F_{1 \Omega}= \frac{\lambda+ e^{-\pi\Omega/a}\lambda\circ
w}{(1- e^{-2\pi\Omega/a})^{{1}/{2}}}$, $F_{2 \Omega}=
\frac{\overline{\lambda\circ w}+
e^{-\pi\Omega/a}\overline{\lambda}}{(1-
e^{-2\pi\Omega/a})^{{1}/{2}}}$. Note that $w(t, x)=(-t, -x)$ is a
wedge reflection isometry that reflects $\lambda$ (defined in the
Rindler region $I$) into $\lambda\circ w$ (in the other region $II$)
\cite{T16,T16A,T16B}.

With the aim of investigating the evolution of the detectors' states
after interacting with the field, the part of the external field
$\phi(f)$ should be traced. Then we obtain final matrix between
Alice and Rob detectors
\begin{eqnarray}
\rho_{t}^{AR} = \left(
                  \begin{array}{cccc}
                    \eta & 0 & 0 & 0 \\
                    0 & 2\mu\sin^2 \theta  & \mu\sin 2\theta  & 0 \\
                    0 & \mu\sin 2\theta  & 2\mu\cos^2 \theta  & 0 \\
                    0 & 0 & 0 & \upsilon \\
                  \end{array}
                \right)
,
\label{rhof2}
\end{eqnarray}
where the parameters $\mu$, $\upsilon$ and $\eta$ respectively take
the form of
\begin{eqnarray}\label{pabc}
\nonumber \mu  &  =\frac{1-q}{2(1-q)+2\nu^{2}(\sin^2 \theta+q\cos^2\theta)},\\
\nonumber \upsilon  &  =\frac{\nu^{2}q\cos^2\theta}{(1-q)+\nu^{2}(\sin^2 \theta+q\cos^2 \theta)},\\
\eta  &  =\frac{\nu^{2}\sin^2\theta}{(1-q)+\nu^{2}(\sin^2
\theta+q\cos^2 \theta)}. \label{abg}
\end{eqnarray}
In addition, the expression of $2\mu+\upsilon+\eta=1$ and the basis
of ${|0_A\rangle\otimes|0_R\rangle},
{|0_A\rangle\otimes|1_R\rangle}$,$ {|1_A\rangle\otimes|0_R\rangle},$
and ${|1_A\rangle\otimes|1_R\rangle}$ are applied in this analysis.
Here the acceleration parameter $a$ is parameterized as $q\equiv
e^{-2\pi\Omega/a}$, and the combined coupling parameter $\nu$
satisfies $\nu^{2}\equiv||\lambda||^{2}=\frac{\epsilon^{2}\Omega\Delta}{2\pi}%
e^{-\Omega^{2}\kappa^{2}}$ \cite{T13,T16}. We remark here that, the
conditions of $\Omega^{-1}\ll\Delta$ and $\nu^2\ll1$ should be
satisfied in the detector model, for the validity of the
perturbation method.

\section{ Quantum Fluctuations of Entanglement}

It's worth noting that, quantum fluctuation determined in terms of
the von Neumann entropy operator is a stochastic quantity, the
fluctuation of which will be taken int account and discussed in this
section. In general, one can use entropy operator $\hat{S}$, or the
so-called entanglement entropy operator $\hat{E}$ (see the ref
\cite{T19,T20} for more details) to quantify the quantum
entanglement. In such case, the entanglement entropy operator
$\hat{E}$ is equivalent to the entropy operator $\hat{S}$, i.e.,
\begin{equation}
\hat{E}=\hat{S}=-\log_2\varrho_A,
\end{equation}
where $\varrho_A=Tr_B(\rho_{AB})$ is the reduced density operator,
with $\rho_{AB}$ denoting the density matrix operator for an
arbitrary pure bipartite system. Hereafter one can define the quantum
fluctuation of entanglement through the entropy operator
\begin{equation}
\Delta{E^2}=\big(\langle\hat{E}^2\rangle-(\langle\hat{E}\rangle)^2\big).
\end{equation}
Following the detailed QFE expression for arbitrary bipartite
systems in both pure and mixed state \cite{T20}, we employ the von
Neumann entropy as a entanglement measurement of a pure bipartite state
\begin{equation}
E=-Tr(\varrho_A \log_2 \varrho_A)=-Tr(\varrho_B \log_2 \varrho_B),
\end{equation}
where $\varrho_B$ represents the reduce density matrix by trace out
subsystem $A$. Actually, the mean value of entropy operator $\hat{E}$ is equivalent to the von Neumann entropy, i.e., $\langle\hat{E}\rangle=E$. For the mixed state with the density matrix in a diagonal form $\xi=\sum_i p_i|\psi_i\rangle \langle\psi_i|$, the definition of the von Neumann entropy changes as follows
\begin{equation}
E=-\sum_i p_i \log_2p_i,
\end{equation}
where $p_i$ is the eigenvalue of the density matrix $\xi$.

Moreover, in the following analysis we also focus on the concurrence
($C$), which has been widely applied in entanglement measurements in
arbitrary bipartite systems \cite{T21}, to quantify quantum
entanglement without statistical fluctuation (since concurrence
doesn't dependent on entropy). Thus, the calculation of the
entanglement fluctuation will reduce to the determination of the
concurrence. Now the fluctuation of quantum entanglement with the
arbitrary bipartite system is quantified as \cite{T22,T23}
\begin{eqnarray}
\Delta E=C \log_2[\frac{1}{C}(1+\sqrt{1-C^2})].
\label{zhang}
\end{eqnarray}
It is worthwhile to note that, different from the case for a pure
state in the bipartite system (with the concurrence of
$C=2\sqrt{\det \varrho_A}=2\sqrt{\det \varrho_B}$), the concurrence
for mixed states takes slightly complicated form. More specifically,
based on the spin flip operation defined by \cite{T21,T24}
\begin{equation}
\tilde{\rho}=(\sigma_y\otimes\sigma_y)\rho^{\ast}(\sigma_y\otimes\sigma_y),
\end{equation}
the concurrence of a bipartite system can be written as
\begin{eqnarray}
\nonumber C&=&\max \{0,\sqrt{\lambda_1}-\sqrt{\lambda_2}-\sqrt{\lambda_3}-\sqrt{\lambda_4}\}, \\
\lambda_i&\geq& \lambda_{i+1}\geq 0.
\label{con}
\end{eqnarray}
where $\lambda_i$ is the eigenvalues with the matrix of $\rho
\tilde{\rho}$.

\begin{figure}[ht]
\centerline{\includegraphics[height=2.25in, width=3.15in]{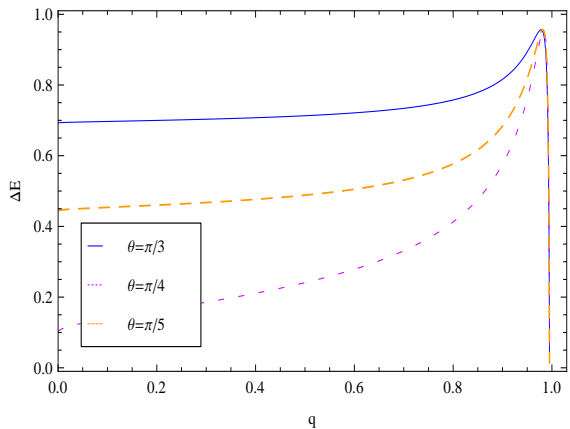}}
\caption{\label{ClaT} The QFE $\Delta E$ as functions of the
acceleration parameter $q$. The initial state entanglement parameter
$\theta$ is fixed at $\theta=\pi/3$ (blue solid line),
$\theta=\pi/4$ (orange dotted line), $\theta=\pi/5$ (violet dotted
line) with the effective coupling parameter $\nu=0.05$.}
\end{figure}

\begin{figure}[ht]
\centerline{\includegraphics[height=2.25in, width=3.15in]{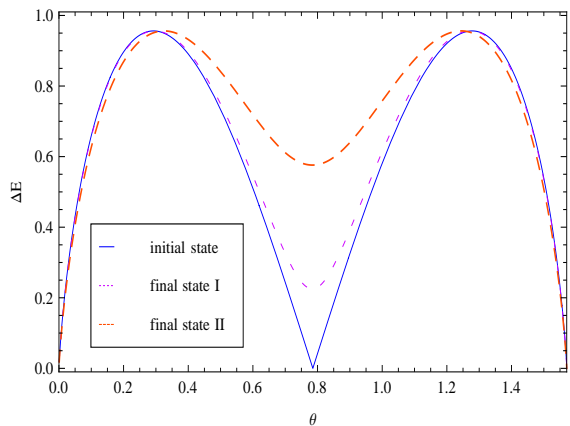}}
\caption{\label{ClaT} The QFE $\Delta E$ of the initial and final
state as a function of the initial entanglement parameter $\theta$.
The effective coupling parameter is fixed at $\nu=0.05$ for final
state I with the $q=0.5$ (violet dotted line) and for final state II
with an extremely large acceleration $q=0.8$ (orange dotted line).
The initial state is denoted as blue solid line.}
\end{figure}

\begin{figure}[ht]
\centerline{\includegraphics[height=2.25in, width=3.15in]{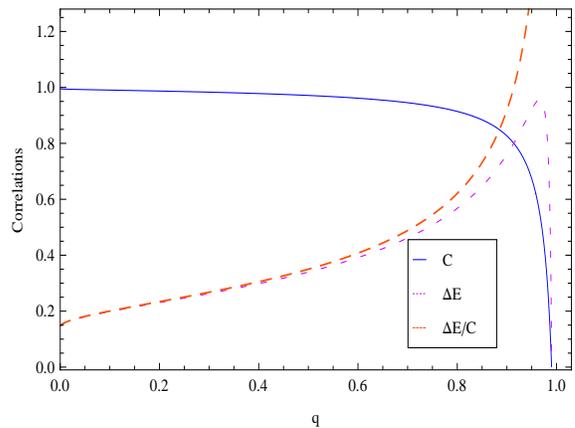}}
\caption{\label{ClaT} The concurrence ($C$), the QFE ($\Delta E$),
and the ratio between concurrence and entanglement ($\Delta E/C$) as
a functions of the detector's acceleration parameter $q$. The
effective coupling parameter is fixed at $\nu=0.05$ for the initial
parameter $\theta=\pi/4$. }
\end{figure}

\section{The behaviors of QFE with the detectors model in a relativistic setting}

In this section we will study the behaviors of QFE under the
influence of the Unruh radiation. With the initial state of the
system given in Eq.~(\ref{IS}), for any nonzero $\theta$, the
detector $A$ shares initial quantum correlation between detector
$R$. Then Rob's detector is accelerated for a time duration $\Delta$
with constant acceleration and influenced by the Unruh thermal bath.
We are interested in how the quantum fluctuations of entanglement is
affected by the relativistic motion of the detectors, specially, the
final mixed state between a pair of Unruh-DeWitt detectors (with the
acceleration of one detector given by Eq.~(\ref{rhof2})). With the
expression of the QFE [Eqs.~(\ref{zhang}) and (\ref{con})], one
could derive the eigenvalues of the $R$ density matrix,
i.e.,
\begin{eqnarray}
\lambda_1=\lambda_2=\eta\upsilon,\,\,\,\,\,
\lambda_3=4\mu^2 \sin^2\theta,\,\,\,\,\,\lambda_4=0,
\end{eqnarray}
where $\mu$, $\upsilon$, and $\eta$ are listed in Eq.~(\ref{abg}).

It is interesting to understand whether the presence of the
detector's acceleration will change the quantum fluctuation of
entanglement (QFE). In Fig.~1, we plot the QFE betweens Alice and
Rob as a function of the detector acceleration parameter $q$, for
three initial state entanglement parameters ($\theta$). The
effective coupling parameter is fixed at $\nu=0.05$. On the one
hand, it is clearly shown that the QFE, which first increases with
the Unruh thermal radiation, will suddenly decays when the
acceleration parameter approaches 0.96. On the other hand, we also
notice here that, as is illustrated in Fig.~1, the initial QFE
($q=0$) is highly dependent on the initial parameter. More
importantly, when the effective coupling parameter is fixed, the
initial QFE of the system (without acceleration effect) will reach
its minimum value at the maximally entangled state and the separable state. For the maximally entangled or the separable state, QFE seems difficult to generate. From the point of the definition of QFE, concurrence has one and zero value corresponding to the separable or the maximally entangled state, respectively, which causes QFE to be zero. Such tendency,
i.e., maximally entangled state will generate less QFE, strongly
indicates the possibility of making use of quantum entanglement to
achieve quantum information tasks. As a final comment, it is
interesting to note that the QFE can not be ignored with the
relativistic motion, because the QFE is very large with extremely
large acceleration, as can be seen from Fig.~1.

Here, we are also interested in the dynamics of QFE for varying
initial parameters, which determines the degree of quantum
entanglement in our analysis ($\theta=\pi/4$ corresponds to the
maximally entangled state). In Fig.~2, we plot the change of the QFE
with the initial state parameter (for fixed effective coupling
parameter $v=0.05$) in the initial state ($q=0$), final state I
($q=0.5$) and final state II ($q=0.9$). Our results show that the
QFE reaches its maximum at the initial parameter $\theta=\pi/8$, $3\pi/8$,
while its minimum value is respectively determined at $\theta=0,
\pi/4, \pi/2$. More interestingly, the oscillatory behavior of QFE
(with a period of $\pi/2$) for different initial parameters is
revealed in this paper. Such tendency can also be seen from the
behavior of $\Delta E$ as a function of $\theta$, in term of
different $q$.

In order to obtain a better understanding of the effect of
statistical fluctuation in entanglement, we illustrate in Fig.~3 the
ratio between the QFE and the concurrence $\Delta E/C$ in terms of
the Rob's acceleration parameter $q$. We emphasize that, the
concurrence can be effectively used to quantify the quantum
entanglement without statistical fluctuation, given the independence
between the concurrence and entropy. By analyzing the behavior of
the concurrence ($C$), the QFE ($\Delta E$), one could clearly see
the effect of the acceleration parameter on the ratio between
concurrence and entanglement. Notice that the ratio of $\Delta E/C$
gradually increases with increasing Rob's acceleration, which
suggests that the Unruh thermal radiation will inevitably makes
quantum entanglement degeneration and concurrently induces quantum
fluctuation of entanglement. In addition, our analysis demonstrates
that although QFE has a huge decay when the acceleration $q$ is
greater than $\sim 0.96$, the ratio of $\Delta E/C$ is still very
large, due to the simultaneous decay of concurrence to a very low
value. Therefore, concerning the realization of out quantum
information task in relativistic setting, one should try to
understand the background mechanism of the loss of quantum
entanglement, and furthermore control the Unruh effect in the case
of lower acceleration.

\vspace*{0.5cm}

\section{Conclusions}

The subject of quantum entanglement continues to be one of great
importance in modern physics. Over the past decades, many of the
studies in this field have concentrated on the realization of
quantum entanglement without considering the effect of acceleration.
However, in realistic situation, the preparation of quantum system
and the procession of quantum information tasks are always
accompanied by accelerated effects. In the framework of such
accelerated quantum system, the Unruh effect will be generated,
which indicates that quantum properties of fields are observer
dependent. Focusing on one of the most general quantum resources,
quantum entanglement, it can be quantified by one statistical
quantity or operator, von Neumann entropy. However, quantum
entanglement has fluctuation under the description of von Neumann
entropy. In this paper, we have investigated the dynamic of quantum
fluctuation of entanglement (QFE) with two entangled Unruh-DeWitt
detectors(modeled by a two-level atom), one of which is accelerated
and interacting with the neighbor external scalar field. Here we
summarize our main conclusions in more detail:

\begin{itemize}

\item Firstly, we find that the QFE initially increases with the Unruh
thermal radiation and then suddenly decays when the acceleration
reaches to $q\sim0.96$, which indicates that QFE can not be ignored
when the relativistic motion is taken into consideration. It is
found that the initial QFE ($q=0$) is highly dependent on the
initial parameter. More specifically, the initial QFE of the system
(without acceleration effect) will reach its minimum value at the
maximally entangled state and the separable
state, which means that maximally entangled
state will bring less QFE. Such findings strongly indicate the
possibility of making use of quantum entanglement to achieve quantum
information tasks.

\item Focusing on the dynamics of QFE for varying
initial parameters, our results show that the QFE reaches its
maximum at the initial parameter $\theta=\pi/8$, $3\pi/8$, while its minimum
value is respectively determined at $\theta=0, \pi/4, \pi/2$. The
investigation of $\Delta E$ as a function of $\theta$ indicates
that, the QFE between Alice and Rob's detectors will exhibit
apparent oscillatory behavior with a period of $\pi/2$, in term of
different value for the initial parameters,

\item In order to quantify the the effect of statistical
fluctuation on entanglement, we employ the ratio between QFE and
quantum entanglement to describe this effect, in which quantum
entanglement is described here in terms of concurrence. Our findings
indicate that the ratio of $\Delta E/C$ gradually increases with
increasing Rob's acceleration, which suggests that the Unruh thermal
radiation will inevitably makes quantum entanglement degeneration
and concurrently induces quantum fluctuation of entanglement. In
addition, our analysis demonstrates that although QFE has a huge
decay when the acceleration $q$ is greater than $\sim 0.96$, the
ratio of $\Delta E/C$ is still very large, due to the simultaneous
decay of concurrence to a very low value.

\item Finally, with the rapid developments in both quantum technology and quantum communication,
it is possible to achieve quantum entanglement by implementing
quantum tasks with relativistic motion in the near future.
Considering the fact that realistic quantum systems always exhibit
gravitational and relativistic features, our analysis in this paper
can be extended to the investigation of the dynamics of QFE, under
the influence of gravitation field. This is supported by General
Relativity, due to the equivalence principle that states all
accelerated reference frames possess a gravitational field.

\end{itemize}
\begin{acknowledgments}
This work is supported by the National Natural Science Foundation
of China under Grant  No. 11675052, No. 11475061, and No. 91536108, the Doctoral Scientific Fund Project of the Ministry of Education of China under Grant No. 20134306120003, and the Postdoctoral Science Foundation of China under Grant No. 2014M560129, No. 2015T80146.

\end{acknowledgments}

\end{document}